\documentclass[twocolumn]{aastex63}

\usepackage{amsmath}
\usepackage{url}

\received{January 20, 2021}
\revised{March 22, 2021}
\accepted{March 22, 2021}
\submitjournal{Astrophysical Journal Letters}

\shorttitle{Reactivation of J1846-0258}
\shortauthors{Blumer et al.}

\graphicspath{{./}{figures/}}

\newcommand\flux{erg\,cm$^{-2}$\,s$^{-1}$}
\newcommand\lum{erg\,s$^{-1}$}
\newcommand\dm{pc~cm$^{-3}$} 
\newcommand\nh{10$^{22}$~cm$^{22}$}
\newcommand\src{PSR~J1846--0258}

\begin{document}

\title{Reactivation of the high-magnetic field pulsar PSR J1846--0258 with magnetar-like bursts}

\correspondingauthor{Harsha Blumer}
\email{harsha.blumer@mail.wvu.edu}

\author{Harsha Blumer}
\affiliation{Department of Physics and Astronomy, West Virginia University, Morgantown, WV 26506, USA}
\affiliation{Center for Gravitational Waves and Cosmology, West Virginia University, Chestnut Ridge Research Building, Morgantown, WV 26505, USA}

\author{Samar Safi-Harb}
\affiliation{Department of Physics and Astronomy, University of Manitoba, Winnipeg, MB R3T 2N2, Canada}

\author{Maura A. McLaughlin}
\affiliation{Department of Physics and Astronomy, West Virginia University, Morgantown, WV 26506, USA}
\affiliation{Center for Gravitational Waves and Cosmology, West Virginia University, Chestnut Ridge Research Building, Morgantown, WV 26505, USA}

\author{William Fiore}
\affiliation{Department of Physics and Astronomy, West Virginia University, Morgantown, WV 26506, USA}
\affiliation{Center for Gravitational Waves and Cosmology, West Virginia University, Chestnut Ridge Research Building, Morgantown, WV 26505, USA}

\begin{abstract}

We report on the 2020 reactivation of the energetic high-magnetic field pulsar \src\ and its pulsar wind nebula (PWN) after 14 years of quiescence with new Chandra and Green Bank Telescope observations. The emission of short-duration bursts from J1846--0258 was accompanied by an enhancement of X-ray persistent flux and significant spectral softening, similar to those observed during its first bursting episode in 2006. The 2020 pulsar spectrum is described by a powerlaw model with a photon index $\Gamma$=1.7$\pm$0.3 in comparison to a $\Gamma$=1.2$\pm$0.1 before outburst, and shows evidence of an emerging thermal component with blackbody temperature $kT$=0.7$\pm$0.1~keV. The 0.5--10 keV unabsorbed flux increased from 5.4$_{-0.5}^{+0.4}\times$10$^{-12}$~\flux \ in quiescence to 1.3$_{-0.2}^{+0.3}\times$10$^{-11}$~\flux \ following the outburst. We did not detect any radio emission from the pulsar at 2 GHz and place an upper limit of 7.1~$\mu$Jy and 55~mJy for the coherent pulsed emission and single-pulses, respectively. The 2020 PWN spectrum, characterized by a photon index of 1.92$\pm$0.04 and X-ray luminosity of (1.2$\pm$0.1)$\times$10$^{35}$~\lum at a distance of 5.8~kpc, is consistent with those observed before the outburst. An analysis of regions closer to the pulsar shows small-scale time variabilities and brightness changes over the 20-yr period from 2000 to 2020, while the photon indices did not change. We conclude that the outburst in \src\ is a combination of crustal and magnetospheric effects, with no significant burst-induced variability in its PWN based on the current observations. 

\end{abstract}


\section{Introduction} 
\label{sec:intro}

\src\ is a high-magnetic field (high-$B$) X-ray pulsar at the center of the supernova remnant (SNR) Kes 75, powering a bright pulsar wind nebula (PWN). It has a rotation period $P$=326 ms, characteristic age $\tau_c$=723 yr, magnetic field $B$=5$\times$10$^{13}$~G, and a spin-down luminosity $\dot{E}$=8.3$\times$10$^{36}$ erg~s$^{-1}$  (Gotthelf et al. 2000). It is the only high-$B$ pulsar with no radio detection reported so far and one of the few pulsars with a measured braking index, with $n$=2.65$\pm$0.01 and spin-down age of 884 yr (Livingstone et al. 2006). \src\ behaved as a typical rotation-powered pulsar for the most part of its observed lifetime, with X-ray luminosity less than its spin-down luminosity.  However, the pulsar underwent a rare event in 2006 when it emitted short magnetar-like bursts detected by Rossi X-ray Timing Explorer (Gavriil et al. 2008), with spectral changes observed with Chandra (Kumar \& Safi-Harb 2008), blurring the distinction between rotation-powered pulsars and magnetars. While in outburst, the pulsar's X-ray luminosity increased by a factor of $\sim$6  and its spectrum softened significantly such that it became reminiscent of those observed from magnetars (Kumar \& Safi-Harb 2008; Ng et al. 2008).

\src\ was observed again in 2009 with Chandra, where its spectrum and flux level were consistent with the quiescent values (Livingstone et al. 2011). The post-outburst timing measurements yielded a braking index of $n$=2.16$\pm$0.13 (and hence, a spin-down age of 1230 yr), suggesting a permanent change in its braking index caused by the magnetar-like burst (Livingstone et al. 2011).  In a followup study with Rossi X-ray Timing Explorer (RXTE) and INTEGRAL, Kuiper \& Hermsen (2009) discovered that the onset of the radiative event was accompanied by a strong glitch in the rotation behaviour of the pulsar. Most recently, an XMM-Newton and NuSTAR study showed that the pulsed emission from \src\ was well-characterized in the 2--50 keV range by a powerlaw (PL) model with photon index $\Gamma$=1.24$\pm$0.09 and a 2--10 keV unabsorbed flux of (2.3$\pm$0.4)$\times$10$^{-12}$~\flux (Gotthelf et al. 2020). There was no evidence for an additional non-thermal component at energies higher than 10 keV, as would be typical for a magnetar. 

On 2020 August 1, the Swift Burst Alert Telescope (BAT) on board the Neil Gehrels Swift Observatory triggered on a soft X-ray burst from the direction of \src\ (Krimm et al. 2020). The BAT data showed a light curve with a single-pulse structure of duration 0.1~s, suggesting the second magnetar-like burst from this high-$B$ pulsar after a quiescent period of 14 years (Laha et al. 2020). The Neutron Star Interior Composition Explorer (NICER) observations reported that the pulsar's pulsed count rate between June-July 2020 in the 2.5--10 keV band was higher compared to normal levels before the detected burst (Kuiper et al. 2020). Radio follow-up observations carried out on August 3, 2020 with the Deep Space Network (DSN) 34-m diameter radio telescope at center frequencies of 8.3 GHz (X-band) and 31.9 GHz (Ka-band) did not find any evidence of radio pulsations from the pulsar (Majid et al. 2020).

In this Letter, we report on an analysis of our Director's Discretionary Time observations of \src\ following its second outburst with Chandra and the Green Bank Telescope, together with all archival Chandra observations from 2000 to 2020.  The distance to \src\ is determined to be roughly 5.8 kpc (Verbiest et al. 2012), as determined through neutral hydrogen absorption, and we scale all derived quantities in units of $d_{5.8}$=$D/5.8$~kpc, where $D$ is the distance. 

\section{Observations}
\label{2}

\subsection{GBT observations}
\label{2.1}

The 100-m Robert C. Byrd Green Bank Telescope (GBT) observed \src\ for 87 minutes at 1.5 GHz (L-band) and for 120 minutes at 2 GHz (S-band) on August 5--6, 2020, for a total exposure of 207 minutes using the Versatile GBT Astronomical Spectrometer (VEGAS) pulsar backend in incoherent dedispersion and total intensity mode. The data were recorded using 512 frequency channels and a sampling time of 81.92~$\mu$s over a bandwidth of 800~MHz. The data were analyzed with the \texttt{PRESTO}\footnote{\url{http://www.cv.nrao.edu/~sransom/presto/}} software package. Radio frequency interference (RFI) excision was performed using the tool \texttt{rfifind}. However, the L-band data were rendered unusable by extremely strong RFI and hence, discarded entirely. We used only the S-band data in this work. 

The S-band data were de-dispersed at various trial dispersion measures (DMs) ranging from 0 to 2000~\dm, as the predicted total Galactic contribution to the DM in this direction is 1470~\dm \ (using the NE2001 model; Cordes \& Lazio 2002). All de-dispersed time series were then folded using the X-ray ephemeris of the pulsar obtained with NICER (Kuiper et al. 2020). No periodicity candidates were found to represent a pulsar signal down to a folded signal-to-noise $S/N$=8.  We also searched each de-dispersed time series for single pulses using the \texttt{PRESTO} tool \texttt{single\_pulse\_search.py}, a boxcar matched-filtering algorithm which compares the data to boxcars in the range of possible widths, up to 100 ms long. The data revealed a nearby source, RRAT J1846--0257, which showed single pulses at a DM around 237~\dm \ (McLaughlin et al. 2006). No other detections were apparent from the single-pulse search. 

With the lack of a detection, we place an upper limit on the coherent flux density of \src\ at 2 GHz, using the following expression from Lorimer \& Kramer (2005):
\begin{equation}
 \label{eq:radiometer1}
    S_\text{min} = \beta\frac{(\text{S/N}_\text{min})T_\text{sys}}{G\sqrt{n_\text{p}t_\text{int}\Delta f}}\sqrt{\frac{W}{P-W}} ~~~\text{mJy},
\end{equation}
where S/N$_\text{min}$ is the minimum signal-to-noise threshold (in this case, 8), $T_\text{sys}$ is the system temperature of the telescope including contribution from background in K, $G$ is the gain of the radio telescope in K/Jy, $n_\text{p}$=2 is the number of polarizations recorded, $t_\text{int}$=7200~s is the integration time of the S-band observation, $\Delta f$=800~MHz is the observing bandwidth, $\beta$=1.1 is a correction factor to account for losses due to digitization, $P$ is the period of \src\, and $W$ is the amount of time per period that the pulsar's emission is ``on", i.e. its pulse width. The GBT observer's guide gives the typical system temperature and gain at this frequency as 18 K and 1.5 K/Jy, respectively. We include a sky background temperature with two components: a 400 K continuum at 408 MHz (Haslam et al. 1982) with an assumed spectral index of $-$2.6, and a contribution from SNR Kes~75 of 10 Jy at 1.4 GHz with a spectral index of $-$0.7 (Green 2006). These values correspond to a flux density of 7.8~Jy for the SNR at 2 GHz. With a duty cycle of 1/64 as in Archibald et al. (2008), our minimum detectable coherent flux density is 7.1~$\mu$Jy. This limit increases for higher duty cycles up to 57~$\mu$Jy for a square wave profile. We can also place an upper limit on the flux density of any single radio pulses at 2 GHz, altering Equation~\ref{eq:radiometer1} to remove the dependence on $P$ and $W$ and changing $t_\text{int}$ from the length of the observation to the length of any hypothetical single pulses. For our single-pulse $S/N$ cutoff of 5, this upper limit is 55~mJy for pulse widths of 3 ms.

\subsection{Chandra observations}
\label{2.2}

\src\ was observed on 2020 September 12 with the Advanced CCD Imaging Spectrometer spectroscopic array (ACIS-S) on board the Chandra X-Ray Observatory for an on-source exposure time of 30 ks (ObsID: 24619). The source was positioned on the back-illuminated S3 chip and the data were taken with a frame time of 0.5~s in VFAINT telemetry format to minimize pileup effects. The standard processing of the data was performed using the \textit{chandra\_repro} script in CIAO version 4.12\footnote{http://cxc.harvard.edu/ciao} (CALDB 4.9.1). The event files were reprocessed from level 1 to level 2 to remove pixel randomization and to correct for CCD charge transfer efficiencies. The bad grades were filtered out and good time intervals were reserved. We have also analyzed all archival observations of \src\ in this study, as in Table~1.
 \begin{deluxetable}{llll}
\tablenum{1}
\tablecaption{Log of Chandra observations of PSR~J1846--0258 \label{tab:logofobs}}
\tablewidth{0pt}
\tablehead{
\colhead{ObsID} & \colhead{Date} & \colhead{Frame} & \colhead{Exposure} \\
& & \colhead{Time} & \colhead{(ks)}
}
\startdata
748 & 15 Oct 2000 & 3.2 & 37.28\\
7337 & 5 Jun 2006 &  1.8 & 17.36\\
6686 & 7 Jun 2006 & 1.8 & 54.07\\
7338 & 9 Jun 2006 & 1.8 & 39.25\\
7339 & 12 Jun 2006 &  1.8 & 44.11\\
10938 & 10 Aug 2009 &  0.4 & 44.12\\
18030 & 8 Jun 2016 &  3.1 & 84.96\\
18866 & 11 Jun 2016 & 3.1 & 61.46\\
24619 &  15 Sept 2020 & 0.5 & 27.78 
\enddata
\tablecomments{Exposure represents the effective exposure time obtained after following the CIAO routines.}
\end{deluxetable}

Figure 1 shows the 0.5--10 keV Chandra image of \src\ ($\alpha_{J2000}$=18$^{h}$46$^{m}$24$^{s}$.94, $\delta_{J2000}$=$-$02$^{\circ}$58\arcmin30\arcsec.1) surrounded by a bright PWN for different epochs, exposure-corrected with a binsize of 1~pixel and smoothed using a Gaussian function of radius 2~pixels. The pulsar appears brighter in 2006 and 2020 following its magnetar-like bursts. The nebula is filled with diffuse emission elongated in the northwest-southeast direction, with prominent northern and southern jets whose morphologies seem to change between different epochs (Ng et al. 2008; Reynolds et al. 2018; Guest \& Safi-Harb 2020).

\begin{figure*}[htp!]
\includegraphics[width=\textwidth]{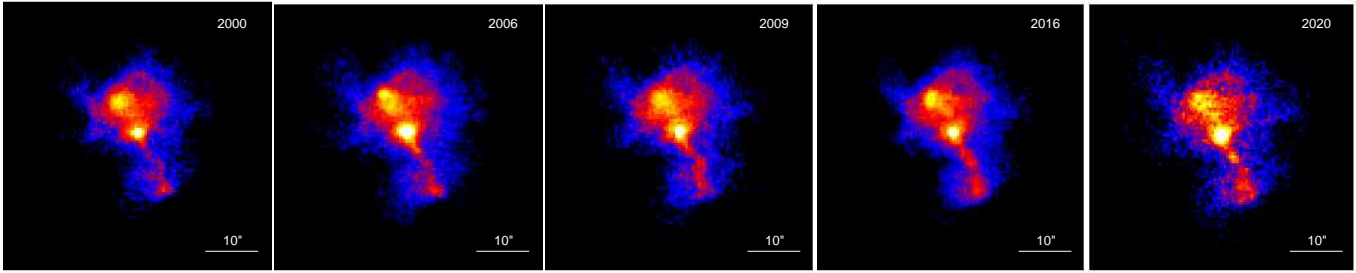}
\caption{Exposure-corrected, logarithmically scaled, images of PSR~J1846--0258 and its bright PWN obtained during different observed epochs (from 2000 to 2020) summarized in Table~1.}
\end{figure*}

\section{Spectroscopy}
\label{3}

The spectral analysis was performed using XSPEC (v12.10.1f) by tying the column density $N_H$ between the epochs to minimize the effects of uncertainties in absorption. The spectra extracted from multiple ObsIDs in the 2006 and 2016 data were combined using the CIAO tool {\it combine\_spectra}  to obtain a merged single spectrum for each of those epochs, because no significant spectral variability was found between the observations.  All the spectra were grouped by a minimum of 30 counts per bin and errors were calculated at the 90\% confidence level. We used the {\it tbabs} absorption model (Wilms et al. 2000) to describe photoelectric absorption by the interstellar medium. As noted by Gotthelf et al. (2020), the spectral fits using the newer column density model resulted in a measured value different from those previously published using the {\it wabs} absorption model (Morrison \& McCammon 1983). 

\begin{figure}[htp!]
\includegraphics[width=0.5\textwidth]{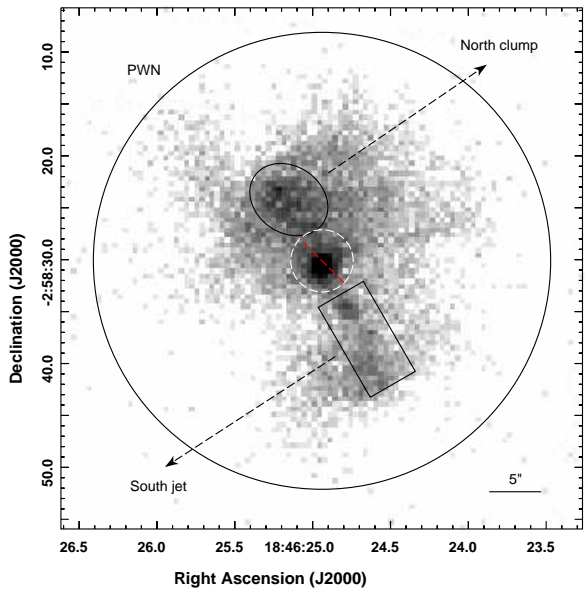}
\caption{Diffuse emission regions used for PWN analysis. PWN: circle excluding the 3$\arcsec$ pulsar region, North clump: small ellipse and south jet: long rectangle. } 
\end{figure}

\subsection{Pulsar spectrum}

The spectrum of \src\ was extracted from a circular region of 1\farcs5 radius, encompassing $\sim$90\% of the encircled energy\footnote{http://cxc.harvard.edu/proposer/POG/html/chap6.html}, and the background from an annular region of 3$\arcsec$--5$\arcsec$ around the source to subtract the contamination by the surrounding PWN. We estimated the impact of photon pileup in all the observations to be $\sim$5--20\% and therefore, a pileup model (Davis et al. 2001) was included during spectral fitting. 

The pulsar spectra for all epochs were simultaneously fit with an absorbed PL which gave the best-fit with a reduced chi-squared value $\chi^2_{\nu}$=1.094 (1207), where $\nu$ is the number of degrees of freedom.  During the 2006 burst, \src\ showed evidence for an additional thermal component in the 0.5--10 keV band (Kumar \& Safi-Harb 2008; Ng et al. 2008). Therefore, the 2020 spectra were further examined by adding a BB component to the model to account for any surface thermal emission from the pulsar. For consistency, we simultaneously fitted the 2006 and 2020 burst data with a BB+PL model, which yielded a $\chi^2_{\nu}$=1.040 (649) over a PL model with $\chi^2_{\nu}$=1.115 (651). An $F$-test indicates that this model is statistically better than a single PL at 99.997\% confidence level. Hence, we conclude that the thermal component is significant in the 2006 and 2020 burst observations. Table~2 lists the spectral parameters and Figure~3 (left) shows the pulsar spectra for all epochs.

\subsection{PWN spectrum}

The PWN spectrum was extracted from a circular region of 22$\arcsec$ radius (excluding the 3$\arcsec$ pulsar region), as in Figure~2, and the background from a 30$\arcsec$ to 40$\arcsec$ annular region centered on the pulsar, to search for any effects of magnetar bursts on the nebula while subtracting the contamination by the surrounding dust scattering halo (see e.g., Gotthelf et al. 2020).  We also extracted regions from near the vicinity of the pulsar, designated as North clump and South jet (Figure~2), and background from adjacent source free elliptical regions to investigate any variability due to the sudden deposition of particles at the time of outburst. The extracted regions for all epochs were simultaneously fit with a PL model and the spectral parameters are shown in Table~2. Figure~3 (right) shows the PWN spectra. The spectral fits were also explored with different backgrounds and binning, and the spectral parameters agree within errors to those shown in Table 2. We refer the readers to Ng et al. (2008), Reynolds et al. (2018), and Guest \& Safi-Harb (2020) for an exhaustive study on the PWN variability and on smaller scales, as we focus here only on the overall burst-induced variability. 

\begin{deluxetable*}{llllllllll}[htp!]
\small
\tablenum{2}
\tablecaption{Spectral fits to \src\ and PWN\label{tab:specfits}}
\tablewidth{0pt}
\tablehead{
\colhead{Epoch} & \colhead{Model} & \colhead{$N_H^{\dag}$} & \colhead{$\Gamma$} & \colhead{$F_{PL}$} & \colhead{$kT$} & \colhead{$F_{BB}$} & \colhead{$L_X$} & \colhead{$L_X/\dot{E}$} \\
& & \colhead{(\nh)} & & \colhead{(\flux)} & \colhead{(keV)} & \colhead{(\flux)} & \colhead{(\lum)} &
}
\startdata
\multicolumn{9}{l}{Pulsar}\\
\cline{1-9}
2000 & PL & 5.7$\pm$0.1 & 1.1$\pm$0.1 & 5.8$_{-0.7}^{+0.8}$$\times$10$^{-12}$ &  &  & 2.3$_{-0.3}^{+0.3}$$\times$10$^{34}$ & 0.0028\\
2006 & PL & 5.7$\pm$0.1 & 1.9$\pm$0.1 & 3.5$_{-0.6}^{+0.5}$$\times$10$^{-11}$ & &   & 1.4$_{-0.2}^{+0.2}$$\times$10$^{35}$ & 0.0169\\
 & BB+PL & 5.6$\pm$0.3 & 1.9$\pm$0.1 & 2.9$_{-0.3}^{+0.2}$$\times$10$^{-11}$ & 0.9$\pm$0.1 & 3.2$_{-0.9}^{+1.2}$$\times$10$^{-12}$ & 1.4$_{-0.3}^{+0.4}$$\times$10$^{35}$ &  0.0169\\
2009 & PL & 5.7$\pm$0.1 & 1.2$\pm$0.1 & 5.7$_{-0.7}^{+0.6}$$\times$10$^{-12}$  & &  & 2.3$_{-0.3}^{+0.3}$$\times$10$^{34}$ & 0.0028 \\
2016 & PL &  5.7$\pm$0.1 & 1.2$\pm$0.1 & 5.4$_{-0.5}^{+0.4}$$\times$10$^{-12}$ &  & & 2.2$_{-0.2}^{+0.2}$$\times$10$^{34}$ & 0.0027 \\
2020 & PL & 5.7$\pm$0.1 & 1.8$\pm$0.1 & 1.3$_{-0.2}^{+0.1}$$\times$10$^{-11}$  & & & 5.2$_{-0.1}^{+0.1}$$\times$10$^{34}$ & 0.0063 \\
 & BB+PL & 5.6$\pm$0.3 & 1.7$\pm$0.3 & 9.2$_{-0.2}^{+0.3}$$\times$10$^{-12}$ & 0.7$\pm$0.1 & 2.3$_{-0.1}^{+0.1}$$\times$10$^{-12}$ & 5.2$_{-0.1}^{+0.2}$$\times$10$^{34}$ & 0.0063 \\[0.35em]
\cline{1-9}
\multicolumn{9}{l}{PWN}\\
\cline{1-9}
2000 & PL & 5.7$\pm$0.1 & 1.93$\pm$0.03 & 3.4$_{-0.2}^{+0.2}$$\times$10$^{-11}$  & & & 1.4$_{-0.1}^{+0.1}$$\times$10$^{35}$ & 0.0169\\
2006 & PL & 5.7$\pm$0.1 & 1.94$\pm$0.02 & 3.6$_{-0.1}^{+0.2}$$\times$10$^{-11}$  & & & 1.5$_{-0.1}^{+0.1}$$\times$10$^{35}$ & 0.0181\\
2009 & PL & 5.7$\pm$0.1 & 1.90$\pm$0.03 & 3.3$_{-0.2}^{+0.2}$$\times$10$^{-11}$  & & & 1.3$_{-0.1}^{+0.1}$$\times$10$^{35}$ & 0.0157 \\
2016 & PL & 5.7$\pm$0.1 & 1.94$\pm$0.02 & 3.2$_{-0.1}^{+0.1}$$\times$10$^{-11}$  & & & 1.3$_{-0.1}^{+0.1}$$\times$10$^{35}$ & 0.0157\\
2020 & PL & 5.7$\pm$0.1 & 1.92$\pm$0.04 & 3.1$_{-0.1}^{+0.1}$$\times$10$^{-11}$  & & & 1.2$_{-0.1}^{+0.1}$$\times$10$^{35}$ & 0.0145\\ [0.35em]
\cline{1-9}
\multicolumn{9}{l}{North clump}\\
\cline{1-9}
2000 & PL & 5.5$\pm$0.1 & 1.77$\pm$0.07 & 7.6$_{-0.6}^{+0.5}$$\times$10$^{-12}$  & & & 3.1$_{-0.2}^{+0.2}$$\times$10$^{34}$ & 0.0037\\
2006 & PL & 5.5$\pm$0.1 & 1.75$\pm$0.04 & 7.9$_{-0.5}^{+0.5}$$\times$10$^{-12}$  & & & 3.2$_{-0.2}^{+0.2}$$\times$10$^{34}$ & 0.0039\\
2009 & PL & 5.5$\pm$0.1 & 1.78$\pm$0.06 & 6.8$_{-0.4}^{+0.5}$$\times$10$^{-12}$  & & & 2.7$_{-0.2}^{+0.2}$$\times$10$^{34}$ & 0.0033\\
2016 & PL & 5.5$\pm$0.1 & 1.77$\pm$0.05 & 6.3$_{-0.5}^{+0.4}$$\times$10$^{-12}$  & & & 2.5$_{-0.2}^{+0.2}$$\times$10$^{34}$ &  0.0030\\
2020 & PL & 5.5$\pm$0.1 & 1.79$\pm$0.09 & 5.9$_{-0.4}^{+0.5}$$\times$10$^{-12}$  & & & 2.4$_{-0.2}^{+0.2}$$\times$10$^{34}$ & 0.0029 \\[0.35em]
\cline{1-9}
\multicolumn{9}{l}{South jet}\\
\cline{1-9}
2000 & PL & 6.0$\pm$0.2 & 1.74$\pm$0.11 & 3.3$_{-0.5}^{+0.5}$$\times$10$^{-12}$  & & & 1.3$_{-0.2}^{+0.2}$$\times$10$^{34}$ & 0.0016\\
2006 & PL & 6.0$\pm$0.2 & 1.77$\pm$0.07 & 3.4$_{-0.3}^{+0.4}$$\times$10$^{-12}$  & & & 1.4$_{-0.2}^{+0.1}$$\times$10$^{34}$ & 0.0017\\
2009 & PL & 6.0$\pm$0.2 & 1.79$\pm$0.10 & 3.2$_{-0.5}^{+0.4}$$\times$10$^{-12}$  & & & 1.3$_{-0.2}^{+0.2}$$\times$10$^{34}$ & 0.0016\\
2016 & PL & 6.0$\pm$0.2 & 1.80$\pm$0.07 & 3.5$_{-0.3}^{+0.4}$$\times$10$^{-12}$  & & & 1.4$_{-0.2}^{+0.1}$$\times$10$^{34}$ & 0.0017\\
2020 & PL & 6.0$\pm$0.2 & 1.77$\pm$0.13 & 3.5$_{-0.5}^{+0.6}$$\times$10$^{-12}$  & & & 1.4$_{-0.2}^{+0.2}$$\times$10$^{34}$ & 0.0017 \\[0.35em]
\enddata
\tablecomments{{\dag}{ -- $tbabs$ XSPEC model with Wilms abundances, tied together in the fit for all epochs.} The uncertainties quoted are 90\% confidence intervals. The unabsorbed fluxes, luminosities, and efficiencies quoted are in the 0.5--10 keV energy range, assuming a distance of 5.8 kpc. The reduced chi-squared values for all extracted regions (models) are Pulsar (PL): $\chi^2_{\nu}$=1.094 (1207), Pulsar (BB+PL): $\chi^2_{\nu}$=1.040 (649), PWN (PL): $\chi^2_{\nu}$=1.093 (1979), North clump (PL): $\chi^2_{\nu}$=1.013 (1056), and South jet (PL): $\chi^2_{\nu}$=0.944 (678).  \\
}
\end{deluxetable*}

\begin{figure*}
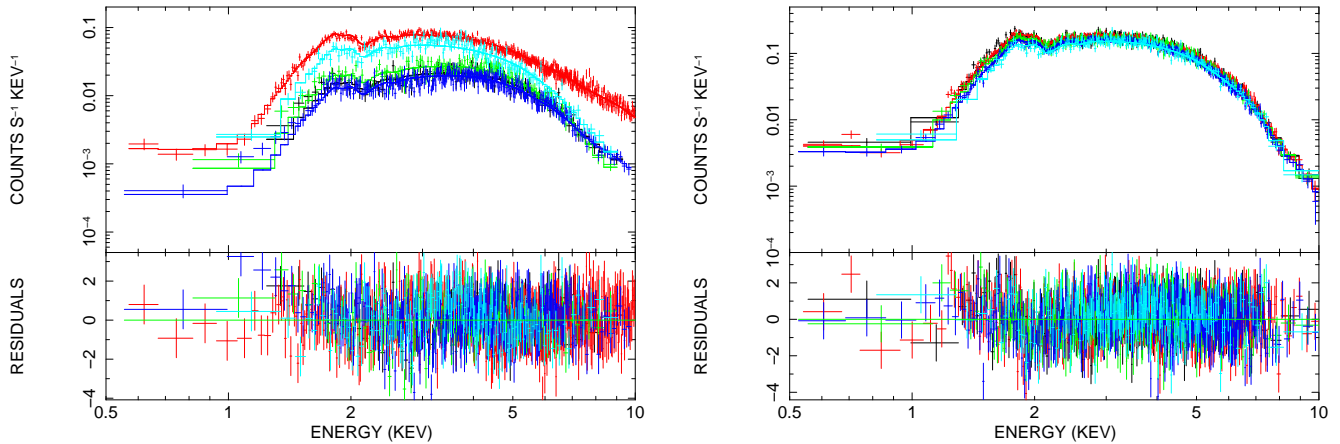

\includegraphics[angle=-90,width=0.5\textwidth]{fig3a.eps}
\includegraphics[angle=-90,width=0.5\textwidth]{fig3b.eps}
\caption{X-ray spectra of \src\ (left) for all epochs. The 2006 (red) and 2020 (cyan) outburst data are fitted by a BB+PL model while the 2000 (green), 2009 (black), and 2016 (blue) quiescent data are fitted by a PL model. The PWN spectra (right) are fitted by a PL model for all epochs. The lower panel shows residuals in units of sigmas with error bars of size one.} 
\end{figure*}

\section{Discussion}

The high-$B$ pulsar \src\ underwent two bursting episodes since its discovery, with the recent outburst in August 2020 after 14 years of quiescence. Radio pulsations have been detected following an X-ray outburst within a few days to less than a year in a few magnetars with a flat spectrum, variable pulse profiles, and flux changes on a timescale from minutes to days over a wide range of frequencies.  However, observations performed with the GBT within 5 days of the burst activity did not result in the detection, with an upper limit of 7.1~$\mu$Jy on the flux at 2~GHz. Radio searches performed 18 months after its X-ray outburst in 2006 also resulted in a non-detection (Archibald et al. 2008).  Radio magnetars are generally characterized by spin-down luminosities larger than their X-ray luminosities, in line with the rotation-powered pulsars, and with $\dot{E}$$\gtrsim$10$^{33}$~\lum, implying an X-ray efficiency $\eta_X$=$L_X/\dot{E}$$<<$1. For PSR~J1846--0258, $\eta_X$$<<$1, but we still have no evidence of radio emission. If the radio emission of \src\ was similar to those of other radio magnetars, the source should have become radio bright sometime after the X-ray outburst. It is possible that the radio emission is beamed out of our line of sight, or is simply too faint. It is unlikely that propagation effects contribute to the non-detection, as the predicted scattering timescale at 2 GHz is only 0.13~ms for the predicted DM of $\sim$318~\dm.

The quiescent X-ray spectrum of \src\ is much like that of other young, high spin-down luminosity pulsars described by a single PL of hard photon index $\Gamma$=1.1$\pm$0.1 and unabsorbed flux $F_{PL}$=5.8$^{+0.8}_{-0.7}$$\times$10$^{-12}$~\flux (Table~2).  During its first magnetar-like bursts in 2006, the pulsar spectrum softened ($\Gamma$=1.9$\pm$0.1) and evolved with the addition of a new thermal component ($kT$=0.9$\pm$0.1~keV) as in Table~2 (see also Kumar \& Safi-Harb 2008; Ng et al. 2008). The 2009 Chandra observations suggested that the pulsar had returned to its quiescence with $\Gamma$=1.2$\pm$0.1. Similar spectral parameters were observed for the 2016 Chandra and 2017 XMM-Newton observations (Gotthelf et al. 2020), suggesting the pulsar has been in quiescence since 2009. Following the latest outburst episode in August 2020, the pulsar spectrum changed again from a single PL to a two-component BB+PL with $\Gamma$=1.7$\pm$0.3 and BB temperature $kT$=0.7$\pm$0.1~keV.  The total unabsorbed flux (0.5--10 keV) was observed to be 1.3$_{-0.2}^{+0.3}\times$10$^{-11}$~\flux, implying an X-ray luminosity $L_{X}$=5.2$_{-0.1}^{+0.2}$$\times$10$^{34}$~$d_{5.8}^2$~erg~s$^{-1}$ and efficiency $\eta_{X}$=$L_{X}/\dot{E}$$\sim$0.0063~$d_{5.8}^2$. According to the twisted-magnetosphere model of magnetars, the crustal fracture may implant twists in the magnetosphere when a crustal event occurs. The twisted fields then induce currents in the magnetosphere and scatters thermal photons to higher energies resulting in a Comptonized BB-like spectrum. Localized hot spots can appear on the star's surface from Ohmic dissipation of the returning currents (Beloborodov 2009) or as a consequence of heat released deeper in the crust by local dissipation of magnetic energy (Pons \& Rea 2012). The neutron star's emitting radius inferred from the BB fit is 1.3$\pm$0.3~km, consistent with the scenario of a hot spot on the star's surface. 
  
It is interesting to compare the present results to those relative to previous outburst activity from \src . The pulsar exhibits very similar characteristics in the new outburst period, confirming the behaviour registered during its past flaring event. The increase in the persistent X-ray flux is ubiquitous in magnetars, often accompanied by hardening of the spectra either in the form of a higher BB temperature or a decrease in the photon index (Kaspi \& Beloborodov 2017). The flux increase in \src\ for the 2020 observations is lower than that observed in 2006, where it showed a flux enhancement by a factor of $\sim$6 times.  It is possible that the pulsar was brighter during the 2020 outburst, but have decreased to its current value at the time of these observations, made 6 weeks from the onset of the burst activity.   Gavriil et al. (2008) reported that the pulsed flux had returned to its quiescent value roughly two months after the 2006 outbursts using RXTE, although no information about the total flux is available from these data. During both bursting episodes, \src\ displayed spectral softening and the emergence of a BB component, which was not detected in the pulsar's quiescent phase. The evolution of the BB component and photon index in \src\ imply that the outburst is a combination of crustal and magnetospheric effects. We also compare \src's outburst properties with the high-$B$ ($\sim$4.1$\times$10$^{13}$~G) radio pulsar PSR~J1119--6127, which has shown a magnetar-like burst in 2016. By contrast, its quiescent spectrum best described by a BB temperature $kT$=0.2$\pm$0.1~keV plus PL $\Gamma$=1.7$\pm$0.7 changed to mainly nonthermal in nature following its 2016 outburst with an index $\Gamma$=2.0$\pm$0.6 (Safi-Harb \& Kumar 2008; Blumer et al. 2017) and exhibited energetically larger flux increase of $>$160 times in the 0.5--10 keV band (Archibald et al. 2016). Regardless, in both pulsars, the inferred X-ray efficiency remained $<<$1 during their peak outburst phase, implying that both pulsars could be powered by their rotational energy, in addition to the magnetic energy that could have caused the outbursts.

The PWN was studied to search for any burst-induced effects on its morphology or spectrum.  Time variabilities are observed for regions near the vicinity of the pulsar, especially the bright clump to the northeast and jet-like feature to the south of the pulsar (Figure~1).  The PWN spectrum in 2020 is well described by a PL with index $\Gamma$=1.92$\pm$0.04, unabsorbed flux of (3.1$\pm$0.1)$\times$10$^{-11}$~\flux, and X-ray luminosity of (1.2$\pm$0.1)$\times$10$^{35}$~$d_{5.8}^2$~\lum in the 0.5--10 keV band. The inferred X-ray efficiency, $\eta_X$=0.0145, is among the highest known for rotation-powered pulsars. These values are consistent within errors to the recent 2017 XMM-Newton observation (Gotthelf et al. 2020), suggesting no significant changes in the overall PWN spectrum. However, the total flux of PWN has decreased by (10$\pm$2)\% over the 20-yr period. The flux appears slightly higher in 2006, possibly due to a dust scattering halo from the magnetar-like burst. The most noticeable change is the decrease in relative brightness of the northern clump by (30$\pm$5)\% over the 20-yr period from 2000 to 2020 while the south jet region has shown a slight increase in brightness, although insignificant within error (Table~2). Interestingly, the photon indices did not change between the epochs for any of our selected regions. While our results agree overall with previous studies (Ng et al. 2008; Reynolds et al. 2018), a direct comparison is not possible due to the different absorption model and scale of regions used. Time-variability of small-scale features are expected in the nebula due to magnetohydrodynamic instabilities in the flow, however, Reynolds et al. (2018) point out that the rapid flux changes between epochs as observed here are difficult to explain with the current models of PWN evolution. 

We conclude that no significant variability was detected in the PWN associated with the magnetar burst for the scales examined. This is not unexpected since the maximum distance the ejected particles from the outburst could have traveled (for $D$=5.8 kpc and $v=c$) is only 1\farcs3. It is possible that at least a fraction of the energetic particles from the pulsar is injected into the regions closer to the pulsar (e.g., southern jet), which could be probed with late-time observations ($>$1 yr from the outburst event). Future high-sensitive monitoring with Chandra would be key to monitor the relaxation and spectral evolution of the pulsar and long-term flux variability in the PWN.   

\acknowledgements
We thank the Green Bank Observatory staff and Chandra Science team for promptly scheduling the observations in Director's Discretionary Time, and Dr. Lucien Kuiper for providing the X-ray ephemeris of PSR J1846--0258. Support for this work is partially provided by the NASA grant DD0-21123X. SSH acknowledges support by NSERC and the Canadian Space Agency. MAM is supported through NSF OIA award number 1458952. WF acknowledges support by the STEM Mountains of Excellence graduate fellowship.


\begin{thebibliography}{}


\bibitem{a08}
Archibald, A. M., Kaspi, V. M., Livingstone, M. A., \& McLaughlin, M. A. 2008, ApJ, 688, 550
\bibitem{a16}
Archibald, R. F., Kaspi, V. M., Tendulkar, S. P., \& Scholz, P. 2016, ApJL, 829, L21
\bibitem{b09}
Beloborodov, A. M. 2009, ApJ, 703, 1044
\bibitem{b17}
Blumer H., Safi-Harb S., \& McLaughlin M. A. 2017, ApJL, 850, L18
\bibitem{cl}
Cordes, J. M., \& Lazio, T. J. W. 2002, arXiv e-prints, astro-ph/0207156

\bibitem{d01}
Davis J. E. 2001, ApJ, 562, 575


\bibitem{}
Gavriil, F. P., Gonzalez, M. E., Gotthelf, E. V., Kaspi, V. M., Livingstone, M. A., \& Woods, P. M. 2008, Science, 319, 1802
\bibitem{}
Green, D. A. 2006, A Catalogue of Supernova Remnants (2006 April version) (Cambridge: Astrophysics Group, Cavendish Laboratory), http://www.mrao.cam.uk /surveys/snrs/
\bibitem{}
Gotthelf, V.E., Vashisht, G., Boylan-Kolchin, M., \& Torri, K.  2000, ApJ, 542L, 37G 
\bibitem{g20}
Gotthelf, E. V., Safi-Harb, S., Straal, S., \& Gelfand, J. 2020, ApJ, submitted
\bibitem{}
Guest, B. T., \& Safi-Harb, S.  2020, MNRAS, 498, 821
\bibitem{}
Haslam, C. G. T., Salter, C. J., Stoffel, H., \& Wilson, W. E. 1982, A\&AS, 47, 1


\bibitem{kb17}
Kaspi, V. M., \& Beloborodov, A. M. 2017, Annu. Rev. Astron. Astrophys, 55, 261
\bibitem{}
Krimm, H. A., Lien, A. Y., Page, K. L., Palmer, D. M. \& Tohuvavohu, A. 2020, GCN Cir. 28187
\bibitem{}
Kumar, H. S., \& Safi-Harb, S. 2008, ApJ, 678, L43
\bibitem{}
Kuiper, L., \& Hermsen, W. 2009, A\&A, 501, 1031
\bibitem{}
Kuiper L., Hermsen W., \& Dekker A., 2018, MNRAS, 475, 1238
\bibitem{}
Kuiper, L., Harding, A., Enoto, T., et al. 2020, ATel 13985
\bibitem{}
Laha, S.;  Barthelmy, S. D.;  Cummings, J. R., et al. 2020, GCN Circ. 28193
\bibitem{}
Livingstone, M. A., Kaspi, V.M., Gotthelf, E.V., \& Kuiper, L. 2006, ApJ, 647, 1286L 
\bibitem{}
Livingstone, M. A., Ng, C.-Y., Kaspi, V., et al. 2011, ApJ, 730, 66
\bibitem{}
Lorimer, D. R., \& Kramer, M. 2005, Handbook of Pulsar Astronomy (Cambridge: Cambridge Univ. Press)
\bibitem{}
Majid, W. A., Pearlman, A. B., Prince, T. A., et al. 2020, ATel, 13988
\bibitem{}
McLaughlin, M. A., et al. 2006, Nature, 439, 817
\bibitem{}
Morrison, R., \& McCammon, D. 1983, ApJ, 270, 119
\bibitem{}
Ng, C. -Y., Slane, P. O., Gaensler, B. M., \& Hughes, J. P. 2008, ApJ, 686, 508



\bibitem{}
Reynolds, S. P., Borkowski, K. J., \& Gwynne, P. H. 2018, ApJ, 856, 133
\bibitem{}
Pons J. A., \& Rea N. 2012, ApJ, 750, L6


\bibitem{}
Verbiest, J. P. W., Weisberg, J. M., Chael, A. A., et al. 2012, ApJ, 755, 39
\bibitem{}
Safi-Harb, S. \& Kumar, H. S. 2008, ApJ, 684, 532


\bibitem{w00}
Wilms J., Allen A., \& McCray R. 2000, ApJ, 542, 914





\end{thebibliography}
\end{document}